\documentclass[twoside,reqno]{HERON}
\usepackage{epsfig,cite,colordvi}
\usepackage{graphicx}
\usepackage{url}
\usepackage{amssymb,amsmath,amscd,epsf}
\usepackage{times}
\usepackage{makeidx}
\usepackage{isotope}
\begin{document}

\title{Islands of shape coexistence in proxy-SU(3) symmetry and in covariant density functional theory}

\runningheads{Islands of Shape Coexistence 
}{D. Bonatsos, K.E. Karakatsanis, A. Martinou, T.J. Mertzimekis, N. Minkov}

\begin{start}

\author{Dennis Bonatsos}{1}, \coauthor{K.E. Karakatsanis}{1,2,3}, \coauthor{Andriana Martinou}{1}, \coauthor{T.J. Mertzimekis}{4}, \coauthor{N. Minkov}{5} 

\index{Bonatsos, D. }
\index{Karakatsanis, K.E.}
\index{Martinou, A.}
\index{Mertzimekis, T.J.}
\index{Minkov, N.} 

\address{Institute of Nuclear and Particle Physics, National Centre for Scientific Research 
``Demokritos'', GR-15310 Aghia Paraskevi, Attiki, Greece}{1}

\address{Department of Physics, Faculty of Science, University of Zagreb, HR-10000 Zagreb, Croatia}{2}

\address{Physics Department, Aristotle University of Thessaloniki, Thessaloniki GR-54124, Greece}{3}

\address{Department of Physics, National and Kapodistrian University of Athens, Zografou Campus, GR-15784 Athens, Greece}{4}

\address{Institute of Nuclear Research and Nuclear Energy, Bulgarian Academy of Sciences, 72 Tzarigrad Road, 1784 Sofia, Bulgaria}{5}

\begin{Abstract}

Shape coexistence in even-even nuclei is observed when the ground state band of a nucleus is accompanied by another K=0 band at similar energy but with radically different structure. We attempt to predict regions of shape coexistence throughout the nuclear chart using the parameter-free proxy-SU(3) symmetry and standard covariant density functional theory. Within the proxy-SU(3) symmetry the interplay of shell model magic numbers, formed by the spin-orbit interaction, and the 3-dimensional isotropic harmonic oscillator magic numbers, leads to the prediction of specific horizontal and vertical stripes on the nuclear chart in which shape coexistence should be possible. Within covariant density functional theory, specific islands on the nuclear chart are found, in which particle-hole excitations leading to shape coexistence are observed. The role played by particle-hole excitations across magic numbers as well as the collapse of magic numbers as deformation sets in is clarified. 

\end{Abstract}
\end{start}

\section{Introduction}

Shape coexistence in atomic nuclei has been receiving attention for a  long time.  Initially proposed in 1956 by Morinaga \cite{Morinaga} in $^{16}$O, it has later been observed in many odd \cite{Meyer} and even-even \cite{Wood,Heyde} atomic nuclei. Efforts in this direction have been recently intensified, since many more experimental examples become available \cite{Garrett} through the advent of radioactive ion beam facilities. 

In the present study we are going to confine ourselves to even-even nuclei, in which shape coexistence appears if the ground state band is accompanied by another $K=0$ band lying nearby in energy, but having radically different structure. For example, one of the two bands can be nearly spherical  and the other one deformed, or both bands can be deformed, but one of them having prolate (rugby ball like) deformation, and the other one possessing oblate (pancake like) deformation. Multiple shape coexistence within the same atomic nucleus has also been reported recently in some cases \cite{GarrettPRL,GarrettPRC}. 

In Fig. 8 of Ref. \cite{Heyde}, the main areas in which experimental evidence of shape coexistence has been observed are schematically indicated by regions resembling ellipses. In medium and heavy mass nuclei such ellipses clearly elongated along the neutron axis in the nuclear chart are seen around the proton magic numbers $Z=82$ and $Z=50$, while ellipses slightly elongated along the proton axis of the nuclear chart are seen around $N=90$ and $N=60$. An additional region is seen around the crossing of $N=40$ and $Z=40$. There are also some smaller regions in lighter nuclei, as well as an elongated region along the $N=Z$ line. 

A more detailed look at the data indicates that even along the $Z=82$ and the $Z=50$ axes the appearance of shape coexistence is not uniform. The Hg $(Z=80$) isotopes are known to be the textbook example \cite{Wood,Heyde} of shape coexistence in the $Z\approx 82$ region. In Fig. 10 of Ref. \cite{Heyde} one sees that shape coexistence is observed only between $N=96$ and $N=110$, in which intruder levels appear among the levels of the ground state band, forming parabolas with a clear minimum at the neutron midshell, $N=104$. Similarly, in Fig. 3.10 of Ref. \cite{Wood}, parabolic curves indicating shape coexistence in the Sn ($Z=50$) isotopes appear only between $N=60$ and $N=70$, again exhibiting a clear minimum at the midshell, $N=64$. These parabolic curves seem to propose that shape coexistence cannot appear anywhere along the neutron axis, but only in regions centered around neutron midshells. 

In relation to the microscopic mechanism being responsible for the appearance of shape coexistence, it has been proposed that particle-hole excitations across proton magic numbers separating major shells in the nuclear shell model are causing the appearance of the effect \cite{Wood,Heyde}. This interpretation appears to hold in the regions along the neutron axis around $Z=82$ and $Z=50$, but it does not offer any answer to the question why shape coexistence appears only around the neutron midshells and not everywhere along the neutron axis. 

In the case of the region around $N=90$, as well as in the very similar region appearing around $N=60$, it has been realized that the particle-hole mechanism across proton magic numbers cannot provide an explanation for the appearance of shape coexistence and that some alternative interpretation is needed (see p. 71 of Ref. \cite{Garrett} and p. 1486 of Ref. \cite{Heyde}, respectively). 

In our recent work we have addressed these questions within two completely different theoretical frameworks, the algebraic proxy-SU(3) scheme \cite{proxy1,proxy2,proxy3} and the microscopic covariant density functional theory approach \cite{Ring1996,Vretenar2005,Niksic2011}. We shall see that the answers provided by these two disparate approaches are fully compatible, indicating in a sense that each approach is corroborating the results of the other. 

 \section{The proxy-SU(3) scheme}
 
The proxy-SU(3) scheme is an approximation which restores the SU(3) symmetry of the Elliott model \cite{Elliott1,Harvey} beyond the $sd$ shell. As it is well known, the nuclear shell model \cite{Mayer1,MJ} is based on the 3-dimensional isotropic harmonic oscillator (3D-HO), bearing the SU(3) symmetry, to which the spin-orbit interaction is added, in order to interpret the nuclear magic numbers 2, 8, 20, 28, 50, 82, 126, \dots, which are different from the magic numbers of the 3D-HO, 2, 8, 20, 40, 70, 112, 168, \dots. In what follows we are going to refer to these two sets as the SM (shell model) and HO (harmonic oscillator) magic numbers, respectively. 

The mechanism through which the spin-orbit interaction is leading from the HO to the SM magic numbers is well known \cite{Mayer1,MJ}. Within each nuclear shell, characterized by ${\cal N}$ quanta, the spin orbit-interaction  affects most strongly the orbital with the highest total angular momentum $j$, which is lowered into the nuclear shell below, having ${\cal N} -1$ quanta. By the same token, the nuclear shell under study, is invaded by the orbital possessing the highest $j$ in the shell above, which has ${\cal N}+1$ quanta. As a result, the shell under study now consists of the remaining orbitals of the ${\cal N}$ shell, plus the intruder orbitals which came from the ${\cal N}+1$ shell above. As a consequence, the SU(3) symmetry of the 3D-HO is destroyed. 

Proxy-SU(3) \cite{proxy1,proxy2,proxy3} restores the SU(3) symmetry, by replacing the intruder orbitals which came from the ${\cal N}+1$ shell by the orbitals of the ${\cal N}$ shell which had escaped into the shell below. It has been shown that these two sets of orbitals bear great similarities. We are going to use the notation of the Nilsson model \cite{Nilsson1,NR} $K [{\cal N}  n_z \Lambda]$, where $\cal{N}$ is the number of oscillator quanta, $n_z$ is the number of oscillator quanta along the $z$-axis, while $\Lambda$ and $K$ are the projections of the orbital angular momentum and the total angular momentum, respectively, along the $z$-axis. In this notation the two sets of orbitals differ by   $\Delta K [\Delta {\cal N} \Delta n_z \Delta \Lambda]$ = 0[110]. We are going to refer to these pairs of orbitals as the 0[110] pairs. It is clear that the 0[110] pairs are characterized by the same projections of orbital angular momentum $\Lambda$, total angular momentum $K$, and spin $\Sigma = K-\Lambda$. In other words, they possess identical rotational properties, their only difference being of vibrational nature, since they differ by one HO quantum. However, this extra quantum lies along the $z$-axis, therefore it does not disturb the cylindrical symmetry of the nucleus \cite{CRC}. 

This replacement is not at all arbitrary. It has been based on the observation that 0[110] pairs are the orbitals for which the proton-neutron interaction exhibits maximum values, as seen empirically through double differences of experimental binding energies \cite{Cakirli}. Furthermore, it has been shown that the 0[110] pairs exhibit maximal spatial overlaps \cite{Karampagia}, a fact that explains the maximization of the interaction, since the nucleon-nucleon interaction is known to be of short range \cite{Castenbook}. In a further step, it has been shown that the particular orbitals forming 0[110] pairs participating in the proxy-SU(3) replacement mentioned above are connected by a unitary transformation, being therefore equivalent \cite{EPJASM}. 

The only dissonance in this harmonic story is that the intruder orbitals  of the ${\cal N}+1$ shell possess one more suborbital than the deserting orbitals of the ${\cal N}$ shell. The extra orbital is the one which possesses the highest projection $m$ of the total angular momentum $j$. However, this orbital, which can accommodate only two alike nucleons anyway, is the one lying highest in energy within its own nuclear shell, therefore it would be empty for most nuclei, not affecting much their structure. This is why Nilsson model calculations before and after the proxy-SU(3) replacement have been found \cite{proxy1} to provide very similar numerical results \cite{proxy1} for the Nilsson diagrams. 

\section{The dual shell mechanism}

The usual magic numbers of the shell model, 2, 8, 20, 28, 50, 82, 126, \dots  are known to be derived from the magic numbers of the three-dimensional isotropic harmonic oscillator (3D-HO), 2, 8, 20, 40, 70, 112, 168, \dots \ by adding  the spin-orbit interaction to the 3D-HO Hamiltonian \cite{Mayer1,MJ}. As already remarked, we  call these two sets the SM (shell model) and HO (harmonic oscillator) magic numbers, respectively. However, if one assumes that the spin-orbit interaction dominates even below 20 particles, one gets the magic numbers 6, 14, 28, 50, 82, 126, \dots \cite{EPJASM}, which we are going to call the SO (spin-orbit) magic numbers. We remark that the SM magic numbers are identical to the HO magic numbers up to 20 particles (the $sd$ shell), while they become identical to the SO magic numbers from 28 particles onward. 

Originating from the isotropic 3D-HO, the SM magic numbers hold for spherical nuclei. For deformed nuclei, one should use an anisotropic harmonic oscillator with cylindrical symmetry (i.e, with two of the three oscillation frequencies along the three cartesian axes being equal), as done in 1955 by Nilsson \cite{Nilsson1,NR}. From the standard Nilsson diagrams, which present the evolution of the energy of the various orbitals with increasing deformation (deviation from sphericity) $\epsilon$, it becomes immediately clear that the SM magic numbers are not valid away from $\epsilon=0$, where the spaghetti of orbitals gets mixed up. This is in accordance with experimental observations of breaking down of various magic numbers as one moves away from the valley of nuclear stability \cite{Sorlin}.

It is expected that the same particle number will correspond in general to a different SU(3) irrep within the SO and HO schemes, since the valence nucleons will be measured from a different magic number (closed shell) up. These irreps can be found in Table 1 of Ref. \cite{EPJASC} for all numbers up to 126.

It is reasonable to assume that away from sphericity the two sets of magic numbers, SO and HO, will play simultaneously complementary roles  in shaping up the properties of the atomic nucleus.  The way this can be done is described in the next section. 

\section{Islands of shape coexistence within the proxy-SU(3) scheme}
  
 A simple SU(3) Hamiltonian reads \cite{Thiamova97,RingSchuck}
 \begin{equation}\label{H0}
 H= H_0- {\kappa\over 2} QQ, \quad \kappa = {\hbar \omega \over 2 N_0}, \quad \hbar \omega = {41 \over A^{1/3}} {\rm MeV}, 
  \end{equation} 
  where $Q$ is the quadrupole operator and $H_0$ is the Hamiltonian of the 3D-HO, its eigenvalues being  
\begin{equation}
N_0 = \sum_{i=1}^A \left({\cal N}_i +{3\over 2} \right), 
\end{equation}
where ${\cal N}_i$ counts the oscillator quanta for each nucleon, given in cartesian and spherical coordinates by 
\begin{equation}
{\cal N} = n_z + n_x +n_y = 2n +l.  
\end{equation} 
  The values of $N_0$ have been calculated and are given up to 126 particles in Table 1 of Ref. \cite{42J}, as well as in Table 1 of Ref. \cite{40J}. 

The quadrupole-quadrupole interaction  $QQ$ is connected to the second order Casimir operator of SU(3) by \cite{DW1,DraayerBook} 
\begin{equation}
QQ = 4 C_2 -3 L(L+1),
\end{equation}  
where $L$ is the eigenvalue of the orbital angular momentum. 
For bandheads of $K=0$ bands in even-even nuclei one has $L=0$. 

For medium-mass and heavy nuclei it is safe to assume that the ground-state band will correspond to the SM irrep, which for these nuclei is the same as the SO irrep, since these irreps have been found to provide very good, {\sl parameter-free} estimates for the collective quantities $\beta$ (deviation from sphericity) and $\gamma$ (deviation from axial symmetry) for a wide range of nuclei \cite{proxy2}, while at the same time they resolve the puzzle of prolate over oblate dominance in the shape of the ground state bands of even-even nuclei \cite{proxy2,proxy3}. Considering the energy difference between the $L=0$ bandheads of the bands built within the SO and HO irreps one finds 
\begin{equation} \label{diff} 
E_{HO}-E_{SO}= (N_{0,HO} -N_{0,SO}) + {\kappa\over 2} (C_{2,SO}-C_{2,HO}), 
\end{equation}   
where $N_{0,HO}$ is given in the above mentioned tables, while the details for the calculation of $N_{0,SO}$ can be found in Ref. \cite{EPJASC}, from which it is seen that for all nuclei one has $(N_{0,HO} -N_{0,SO})\leq 0$.  One also has $(E_{HO}-E_{SO})\geq 0$, since SO corresponds to the ground state band. Therefore, for Eq. (\ref{diff}) to be valid,  one should have $(C_{2,SO}-C_{2,HO})\geq 0$, since $\kappa \geq 0$ from Eq. (\ref{H0}).  
One can easily see \cite{EPJASC} that the condition  $C_{2,SO} \geq C_{2,HO}$ is fulfilled within the regions of 7-8, 17-20, 34-40, 59-70, 96-112, 146-168 nucleons.  
   
Let us discuss the physical consequences of this condition. In order to have two bands exhibiting shape coexistence, they should lie close in energy, i.e. $E_{HO}-E_{SO}$ should be positive, but close to zero. Then from Eq. (\ref{diff}) one should have $C_{2,SO}-C_{2,HO} \geq N_{SO}-N_{HO} \geq 0$, which implies that the shape coexistence region will start at the point at which $C_{2,SO} \approx C_{2,HO}$. In other words, shape coexistence starts when the eigenvalue of the second order Casimir of SU(3) for the SO irrep starts exceeding  that of the HO irrep, while it ends where the HO magic number is reached \cite{EPJASC}. 
  
\begin{figure}[t]
    \centering
    \includegraphics[scale=0.1]{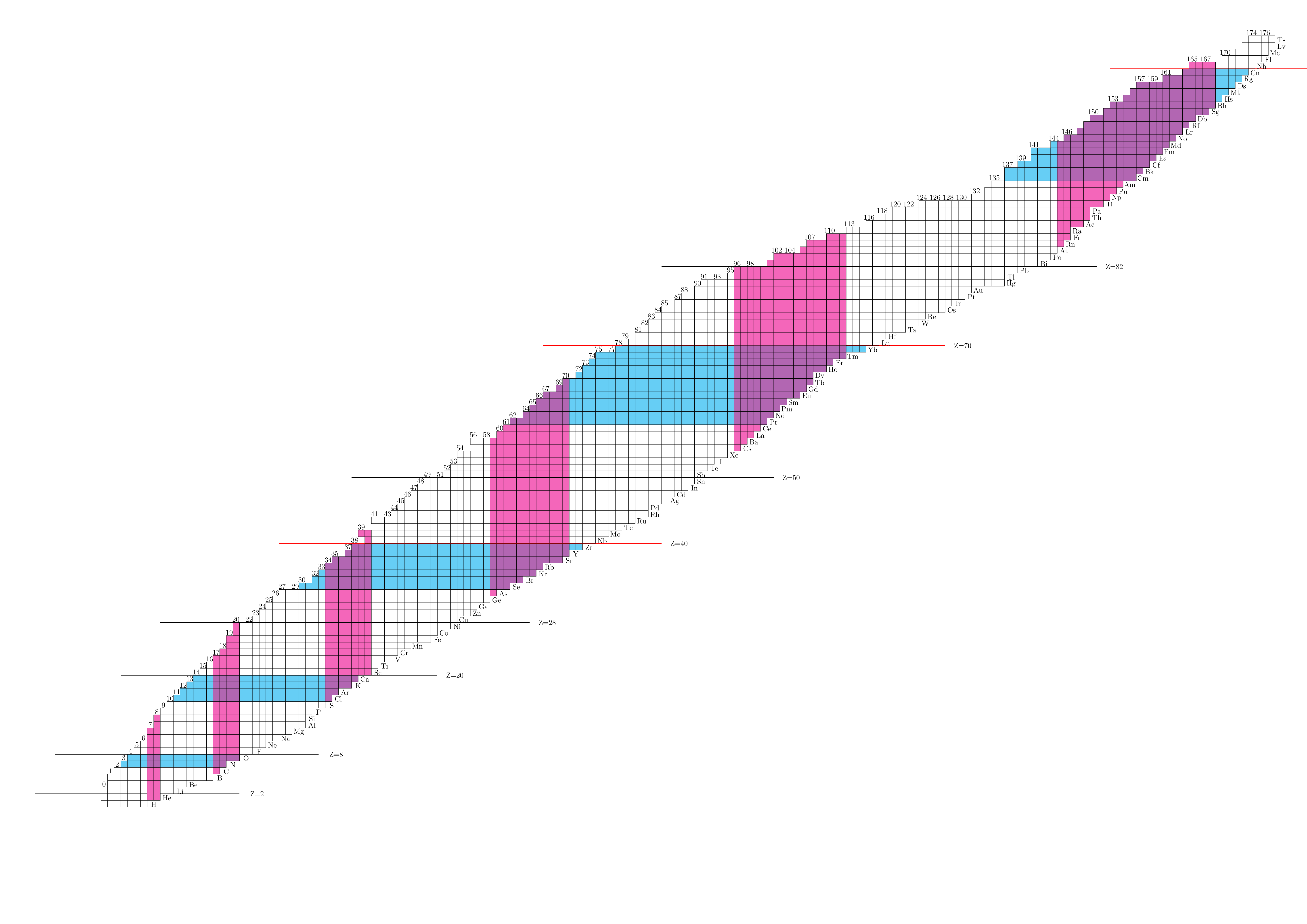}
    \caption{The colored regions in this map possess proton or neutron number between 7-8, 17-20, 34-40, 59-70, 96-112, 145-168. The horizontal stripes correspond to the proton induced shape coexistence, while the vertical stripes correspond to the neutron induced shape coexistence. Therefore the map indicates, which nuclei have to be examined both theoretically and experimentally for manifesting shape coexistence according to the proposed mechanism. Adopted from Ref. \cite{EPJASC}.}
    \label{map}
\end{figure}

The regions of the nuclear chart in which shape coexistence can occur are depicted in Fig.\,1. It should be noticed that the red and blue stripes shown there indicate the widest possible regions within which shape coexistence could appear. In other words, they represent a necessary, but not sufficient condition for shape coexistence. An effort to narrow down these stripes into islands through a necessary condition will be undertaken in the next section.   

 \section{Islands of shape coexistence within covariant density functional theory} 
 
Covariant density functional theory has become an indispensable tool for microscopic predictions of basic nuclear properties \cite{Ring1996,Vretenar2005,Niksic2011}, as well as for the microscopic derivation of model parameters needed within algebraic collective models, like the Interacting Boson Model \cite{IA}, from which detailed predictions for spectra and electromagnetic transition rates can be obtained \cite{Nomura1,Nomura2}. 

We have recently attempted \cite{PLBSC,PRCSC} to search for the microscopic roots of shape coexistence by calculating the single particle energy levels of protons and neutrons in several nuclei and looking for proton or neutron particle-hole excitations predicted there. Standard covariant density functional theory has been used for our calculations. In particular, the DDME2 functional of Ref. \cite{Lalazissis} is used within the code of Ref. \cite{Niksic}. The single particle energies are determined, labeled by Nilsson quantum numbers \cite{NR}, using the method described in Refs. \cite{Prassa,Karakatsanis1,Karakatsanis2}. 
 
\begin{table}[b]
\caption{Proton single-particle energy levels participating in proton particle-hole formation in various isotopes across different regions of the nuclear chart.  Since the proton particle-hole excitations are caused by the neutrons, we say that neutron-induced shape coexistence is expected in these isotopes. See Section 5 for further discussion. Adopted from \cite{PRCSC}. }
\smallskip
\small\noindent\tabcolsep=9pt

\begin{tabular}{ l  l  l  }

\hline
nuclei                           & occupied $Z>40$      & vacant $Z<40$ \\ 
\hline 
\medskip
\isotope[78][38]Sr          &  1/2[440] 3/2[431]          &   3/2[301] 5/2[303]     \\ \medskip
\isotope[78,80][40]Zr    &  1/2[440] 3/2[431] 5/2[422] &   1/2[301]  3/2[301] 5/2[303]     \\ 

\hline
nuclei                          & occupied $Z>50$      & vacant $Z<50$ \\ 
\hline 
\medskip
\isotope[116-120][52]Te   &      3/2[422]     &    9/2[404]   \\ 

\hline
nuclei                           & occupied $Z>82$      & vacant $Z<82$ \\ 
\hline 
\medskip

\isotope[176,188][78]Pt   & 1/2[541]             &                1/2[400]  3/2[402] \\ \medskip
\isotope[178-186][78]Pt  & 1/2[541] 3/2[532]   &                 1/2[400]  3/2[402]    \\ \medskip
\isotope[176][80]Hg           & 1/2[541]             &  11/2[505]     1/2[400]  3/2[402]  \\ \medskip
\isotope[178-190][80]Hg   & 1/2[541] 3/2[532]   &   11/2[505]     1/2[400]  3/2[402] \\ \medskip
\isotope[180-190][82]Pb   & 1/2[541] 3/2[532]   &   11/2[505]     1/2[400]  3/2[402]  \\ \medskip 
\isotope[182-192][84]Po   & 1/2[541] 3/2[532]   &   11/2[505]     1/2[400]              \\ 
\hline

\end{tabular}
\end{table} 

\begin{table}[b]
\caption{Neutron single-particle energy levels participating in neutron particle--hole formation in various isotones across different regions of the nuclear chart.  Since the neutron particle-hole excitations are caused by the protons, we say that proton-induced shape coexistence is expected in these isotones. See Section 5 for further discussion. Adopted from \cite{PRCSC}. }
\smallskip
\small\noindent\tabcolsep=9pt

\begin{tabular}{ l  l  l  }

\hline
nuclei                           & occupied $N>40$      & vacant $N<40$ \\ 
\hline 
\medskip
\isotope[78][40]Zr                            &  1/2[440] 3/2[431]          &    3/2[301]  5/2[303]            \\ \medskip
\isotope[78][38]Sr, \isotope[80][40]Zr     &  1/2[440] 3/2[431] 5/2[422] &  1/2[301]  3/2[301] 5/2[303]     \\

\hline
nuclei                           & occupied $N>70$      & vacant $N<70$ \\ 
\hline 
\medskip
\isotope[98][40]Zr                              &  1/2[550]           & 1/2[411]  5/2[413]       \\ \medskip
\isotope[100][40]Zr, \isotope[102][42]Mo     &  1/2[550] 3/2[541]  & 1/2[411]  5/2[413]       \\

\hline
nuclei                           & occupied $N>112$      & vacant $N<112$ \\ 
\hline 
\medskip

\isotope[150][60]Nd,  \isotope[152][62]Sm, \isotope[154][64]Gd & 1/2[660] &   5/2[523]    \\ \medskip
\isotope[152][60]Nd,  \isotope[154][62]Sm, \isotope[156][64]Gd & 3/2[651] &   5/2[523]    \\ \medskip 

\end{tabular}
\end{table}

The results obtained so far are summarized in Tables 1 and 2, and in Fig.\,2, to which several comments apply. 

In Table\,1 are listed the nuclei in which {\sl proton} particle-hole excitations have been found \cite{PRCSC}. They are clustered into a large island around the magic number $Z=82$, as well as into a smaller island in the region of the magic number $Z=50$. These islands are shown in yellow color in Fig.\,2, and correspond to islands L and J respectively in Fig.\,8 of Ref. \cite{Heyde}.  It is clear that within each series of isotopes, particle-hole excitations appear around the midshells $N=104$ and $N=66$. Since in each series of isotopes the same protons are present, it is reasonable to conclude that the p-h excitations are due to the proton-neutron interaction, which should be stronger around mid-shell, since more valence neutrons are present there. Therefore we refer to this case as {\sl neutron-induced} shape coexistence.

 \begin{figure}[htb]
\centering
\includegraphics[width=110mm]{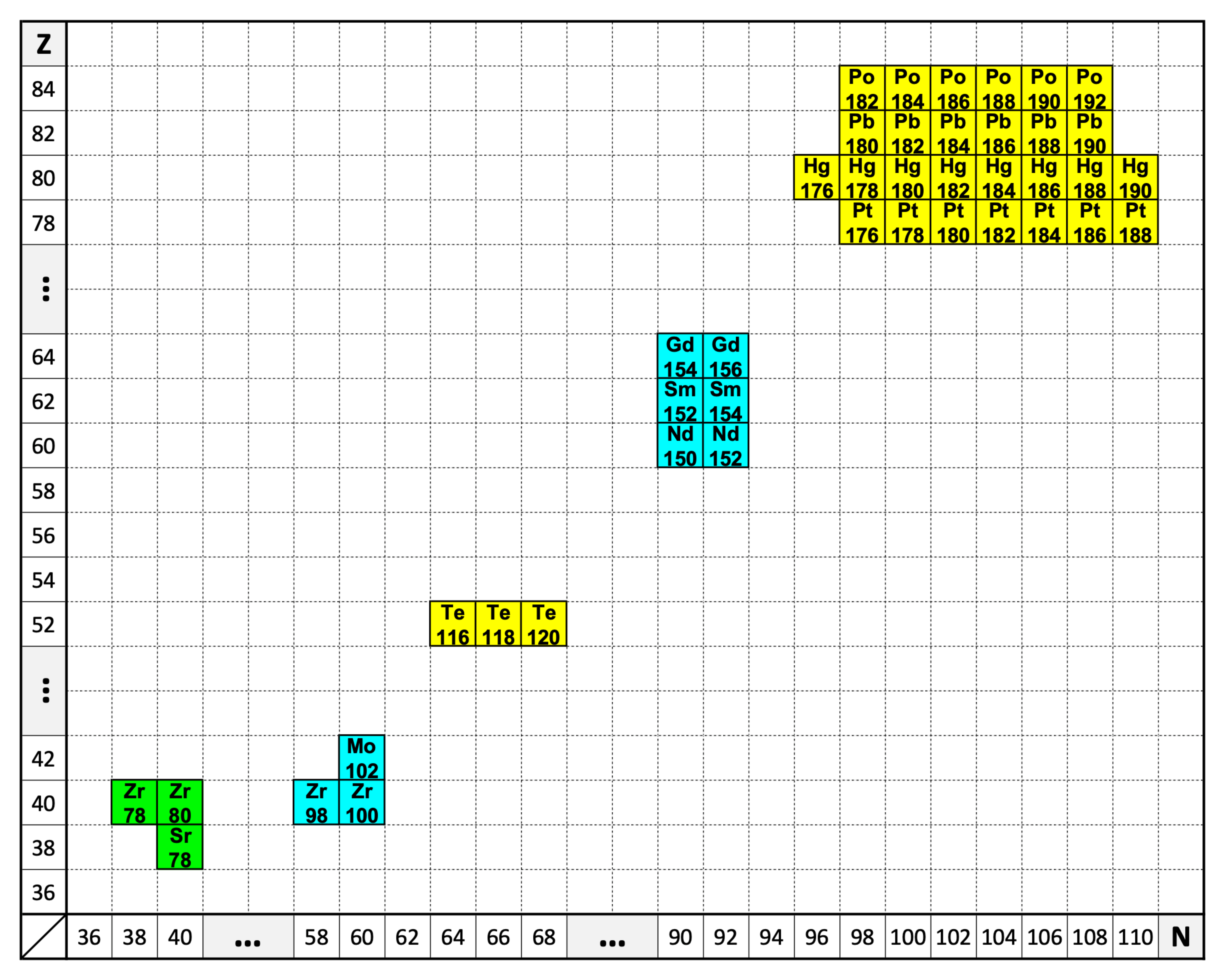}
\caption{Islands of shape coexistence (SC) found through covariant density functional theory calculations \cite{PLBSC,PRCSC}. Islands corresponding to neutron-induced SC are shown in yellow, islands due to proton-induced SC are exhibited in cyan, while islands due to both mechanisms are shown in green. See Section 5  for further discussion. Adopted from \cite{PRCSC}.}
\end{figure}

In Table\,2 are listed the nuclei in which {\sl neutron} particle-hole excitations have been found \cite{PRCSC}. They are clustered into a larger island around $N=90$, as well as into a smaller island around  $N=60$. These islands are shown in cyan color in Fig.\,2, and correspond to islands K and I respectively in Fig.\,8 of Ref. \cite{Heyde}. It is clear that within each series of isotones, particle-hole excitations appear close to the midshells $Z=66$ and $Z=39$. Since in each series of isotones the same neutrons are present, it is reasonable to conclude that the p-h excitations are due to the proton-neutron interaction, which should be stronger around mid-shell, since more valence protons are present there. Therefore we refer to this case as {\sl proton-induced} shape coexistence. 

Furthermore in both Tables 1 and 2 are listed nuclei in which {\sl both} proton and neutron particle-hole excitations appear. They are clustered into an island around $Z=N=40$, shown in green color in Fig.\,2, which corresponds to island H  in Fig.\,8 of Ref. \cite{Heyde}. In this case, both protons and neutrons lie close to the $Z=N=39$ midshells, therefore shape coexistence in this case is both proton-induced and neutron-induced. 

The islands of Fig.\,2 do lie within the stripes of Fig.\,1, thus offering a hint on how the stripes of Fig.\,1 can be narrowed down into rounded islands. Therefore the appearance of p-h excitations can be considered as a sufficient condition for the appearance of shape coexistence. However, a word of warning applies here. The islands of Fig.\,2 correspond to the cases in which the proton-neutron interaction is strong enough in order to create p-h excitations in the ground state configuration of the nucleus. There might be additional cases, in which p-h excitations are created in higher configurations, despite the fact that no p-h excitations are induced in the ground state configuration. This might indicate that the sufficient condition depicted in Fig.\,2 might be too restrictive and that the islands of shape coexistence which could appear within the stripes of Fig.\,1 might be larger than those seen in Fig.\,2. This is a call for further investigations.

\section{Conclusion}

Our investigations on shape coexistence, both through parameter-independent arguments within the proxy-SU(3) scheme and with standard covariant density functional theory calculations converge to the conclusion that shape coexistence cannot appear all over the nuclear chart, but it is confined within certain islands, within which the proton-neutron interaction is strong enough to create particle-hole excitations. These islands emerge through an interplay of the usual 3D-HO (HO) magic numbers, which are valid in the absence of any spin-orbit interaction, and the spin-orbit (SO) magic numbers, which prevail when the spin-orbit interaction is strong. The standard magic numbers of the shell model, which  are a mixture of HO and SO magic numbers, with the former prevailing in light nuclei up to the $sd$ shell, and the latter dominating in heavier nuclei, are valid only at zero deformation. With deformation gradually setting in, the interplay of HO and SO magic numbers prevails. Since this interplay manifests itself both in protons and in neutrons, four different proton-neutron cases might occur (SO-SO, SO-HO, HO-SO, HO-HO), which should be further investigated, since they might lead to concrete predictions of multiple coexistence.   
 
\section*{Acknowledgements} 

Support by the Tenure Track Pilot Programme of the Croatian Science Foundation and the Ecole Polytechnique F\'{e}d\'{e}rale de Lausanne, the Project TTP-2018-07-3554 Exotic Nuclear Structure and Dynamics with funds of the Croatian-Swiss Research Programme, as well as by the Bulgarian National Science Fund (BNSF) under Contract No. KP-06-N48/1  is gratefully acknowledged, with parallel lack of any support by the Hellenic Foundation for Research and Innovation (HFRI) pointed out. Results presented in Section 5 have been produced using the Aristotle University of Thessaloniki (AUTH) computational infrastructure and resources. The authors would like to acknowledge the support provided by the IT Center of AUTH throughout the progress of this research work. 








\begin{thebibliography}{99}

\bibitem{Morinaga}
H. Morinaga, \textit{Phys. Rev.} \textbf{101}(1956) 254. 

\bibitem{Meyer}
K. Heyde, P. Van Isacker, M. Waroquier, J.L. Wood, and R.A. Meyer, \textit{Phys. Rep.} \textbf{102} (1983) 291.  

\bibitem{Wood}
J.L. Wood, K. Heyde, W. Nazarewicz, M. Huyse, and P. Van Duppen, \textit{Phys. Rep.} \textbf{215}(1992) 101. 

\bibitem{Heyde}
K. Heyde and J.L. Wood, \textit{Rev. Mod. Phys.} \textbf{83} (2011) 1467. 

\bibitem{Garrett}
P.E. Garrett, M. Zieli\'nska, and E. Cl\'ement, \textit{Prog. Part. Nucl. Phys.} \textbf{124} (2022) 103931. 

\bibitem{GarrettPRL}
P.E. Garrett {\it et al.}, Phys. Rev. Lett. \textbf{123}(2019) 142502. 

\bibitem{GarrettPRC}
P.E. Garrett {\it et al.}, Phys. Rev. C \textbf{101} (2020) 044302. 

\bibitem{proxy1}
D. Bonatsos, I. E. Assimakis, N. Minkov, A. Martinou, R. B. Cakirli, R. F. Casten,  and K. Blaum, \textit{Phys. Rev. C}   \textbf{95} (2017) 064325. 

\bibitem{proxy2}
D. Bonatsos, I. E. Assimakis, N. Minkov, A. Martinou, S. Sarantopoulou, R. B. Cakirli, R. F. Casten, and K. Blaum, \textit{Phys. Rev. C} \textbf{95} (2017) 064326.

\bibitem{proxy3}
D. Bonatsos,  \textit{Eur. Phys. J. A} \textbf{53} (2017) 148.

\bibitem{Ring1996}
P. Ring, \textit{Prog. Part. Nucl. Phys.} \textbf{37} (1996) 193.

\bibitem{Vretenar2005}
D. Vretenar, A.V. Afanasjev, G.A. Lalazissis and P. Ring, \textit{Phys. Rep.} \textbf{409} (2005) 101. 

\bibitem{Niksic2011}
T. Nik\v{s}i\'{c}, D. Vretenar, and P. Ring, \textit{Prog. Part. Nucl. Phys.} \textbf{66} (2011) 519.

\bibitem{Elliott1}
J.P. Elliott, \textit{Proc. Roy. Soc. London Ser. A: Math. Phys. Sci.} \textbf{245} (1958) 128. 

\bibitem{Harvey}
M. Harvey, \textit{The Nuclear SU(3) Model}, in \textit{Advances in Nuclear Physics} \textbf{1}, ed. M. Baranger and E. Vogt, Prenum, New York (1968) p. 67.  

\bibitem{Mayer1}
M. G. Mayer,  \textit{Phys. Rev.} \textbf{74} (1948) 235. 

\bibitem{MJ}
M.G. Mayer and J.H.D. Jensen, {\it Elementary Theory of Nuclear Shell Structure}, Wiley, New York (1955). 

\bibitem{Nilsson1} 
S. G. Nilsson, \textit{Mat. Fys. Medd. K. Dan. Vidensk. Selsk.} \textbf{29} (1955) no. 16.

\bibitem{NR}
S. G. Nilsson and I. Ragnarsson, \textit{Shapes and Shells in Nuclear Structure}, Cambridge University Press, Cambridge (1995). 

\bibitem{CRC}                                                              
A. Martinou and D. Bonatsos, in \textit{Nuclear Structure Physics}, ed. A. Shukla and S.K. Patra, CRC Press, Boca Raton, (2020) 1. 

\bibitem{Cakirli}
R.B. Cakirli, K. Blaum, and R.F. Casten, \textit{Phys. Rev. C} \textbf{82} (2010) 061304(R).

\bibitem{Karampagia}
D. Bonatsos, S. Karampagia, R.B. Cakirli, R.F. Casten, K. Blaum,  and L. Amon Susam,  \textit{Phys. Rev. C} \textbf{88} (2013) 054309.

\bibitem{Castenbook}
R. F. Casten, {\it Nuclear Structure from a Simple Perspective}, Oxford University Press, Oxford (1990).

\bibitem{EPJASM}
A. Martinou, D. Bonatsos, N. Minkov, I.E. Assimakis, S.K. Peroulis, S. Sarantopoulou, and J. Cseh, \textit{Eur. Phys. J. A} \textbf{56} (2020) 239. 

\bibitem{Sorlin}
O. Sorlin and M.-G. Porquet, \textit{Prog. Part. Nucl. Phys.} \textbf{61} (2008) 602. 

\bibitem{EPJASC}
A. Martinou, D. Bonatsos, T.J. Mertzimekis, K. Karakatsanis, I.E. Assimakis, S.K. Peroulis, S. Sarantopoulou, and N. Minkov, \textit{Eur. Phys. J. A}  \textbf{57} (2021) 84. 

\bibitem{Thiamova97} 
D. J. Rowe, G. Thiamova, and J. L. Wood, \textit{Phys. Rev. Lett.}  \textbf{97} (2006) 202501.

\bibitem{RingSchuck}
P. Ring and P. Schuck, {\it The Nuclear Many Body Problem}, Springer-Verlag,  Berlin (1980).

\bibitem{42J}                                                       
A. Martinou, S. Sarantopoulou, K.E. Karakatsanis, and D. Bonatsos, \textit{Eur. Phys. J. Web of Conferences} 252 (2021) 02006.  

\bibitem{40J}
S. Sarantopoulou, A. Martinou, and D. Bonatsos, \textit{Bulg. J. Phys.} \textbf{46} (2019) 455. 

\bibitem{DW1}
J. P. Draayer and K. J. Weeks, \textit{Phys. Rev. Lett.} \textbf{51} (1983) 1422.

\bibitem{DraayerBook}
J.P. Draayer, in \textit{Algebraic Approaches to Nuclear Structure: Interacting boson and fermion models}, ed. R.F. Casten, Harwood, Chur  (1993) p. 423. 

\bibitem{IA}
F. Iachello and A. Arima, {\it The Interacting Boson Model}, Cambridge University Press, Cambridge (1987). 

\bibitem{Nomura1}
K. Nomura, N. Shimizu, and T. Otsuka, \textit{Phys. Rev. Lett.} \textbf{101} (2008) 142501.

\bibitem{Nomura2}
K. Nomura, N. Shimizu, and T. Otsuka, \textit{Phys. Rev. C }\textbf{81} (2010) 044307.

\bibitem{PLBSC}
D. Bonatsos,  K.E. Karakatsanis, A. Martinou, T.J. Mertzimekis,  and N. Minkov, \textit{Phys. Lett. B} \textbf{829} (2022) 137099.

\bibitem{PRCSC}
D. Bonatsos,  K.E. Karakatsanis, A. Martinou, T.J. Mertzimekis,  and N. Minkov, to be published.

\bibitem{Lalazissis}
G.A. Lalazissis, T. Nik\v{s}i\'{c}, N. Paar, D. Vretenar, and P. Ring, \textit{Phys. Rev. C} \textbf{71} (2005) 024312.

\bibitem{Niksic}
T. Nik\v{s}i\'{c}, N. Paar, D. Vretenar, and P. Ring, \textit{Comp. Phys. Commun.} \textbf{185} (2014) 1808. 

\bibitem{Prassa}
V. Prassa, T. Nik\v{s}i\'{c}, and D. Vretenar, \textit{Phys. Rev. C} \textbf{88} (2013) 044324. 

\bibitem{Karakatsanis1}
K.E. Karakatsanis, Ph.D. thesis, U. Thessaloniki (2017).\hfill\break  http://hdl.handle.net/10442/hedi/41280. 

\bibitem{Karakatsanis2}
K.E. Karakatsanis, G.A. Lalazissis, V. Prassa, and P. Ring, \textit{Phys. Rev. C} \textbf{102} (2020) 034311. 

\end{thebibliography}
\end{document}